\documentclass[]{article}
\usepackage[margin=1in]{geometry}
\usepackage{setspace, graphicx, amsmath}
\usepackage{amssymb}
\usepackage{verbatim} 
\usepackage{booktabs}
\usepackage{dirtytalk}
\usepackage{float}
\usepackage{graphicx}
\usepackage{subcaption}
\usepackage{mwe}

\newtheorem{prop}{Proposition}[section]

\graphicspath{{./graph/}}

\usepackage{color}

\def\boxit#1{\vbox{\hrule\hbox{\vrule\kern6pt\vbox{\kern6pt#1\kern6pt}\kern6pt\vrule}\hrule}}

\title{Method of Moments Confidence Intervals for a Semi-Supervised Two-Component Mixture Model}
\author{Bradley Lubich, Daniel Jeske, Weixin Yao}
\date{}
\begin{document}

\maketitle

\begin{abstract}
	A mixture of a distribution of responses from untreated patients and a shift of that distribution is a useful model for the responses from a group of treated patients. The mixture model accounts for the fact that not all the patients in the treated group will respond to the treatment and consequently their responses follow the same distribution as the responses from untreated patients. The treatment effect in this context consists of both the fraction of the treated patients that are responders and the magnitude of the shift in the distribution for the responders. In this paper, we investigate properties of the method of moment estimators for the treatment effect and demonstrate their usefulness for obtaining approximate confidence intervals without any parametric assumptions about the distribution of responses.
\end{abstract}

\section{Introduction}
	In a control group vs$.$ treatment group design, mixture models$^{1,2}$ can be a good choice for the treatment group response distribution in anticipation that there might be a sub-population of the treated population whose responses still come from the control group distribution. It is well known that such sub-populations of `non-responding' treated patients exist in oncology trials$^{3,4,5}$. A group fMRI example motivated a recent call for more attention to be given to mixture alternatives for comparing two (alternative) treatments (by a hypothesis test), stating that medical applications, psychiatric-genetics and personalized medicine are important applications where mixtures are plausible alternatives$^6$. Jeske and Yao$^7$ demonstrated that ignoring the heterogeneity of treatment effects could result in an under-powered experiment and have the risk of missing some useful treatments. When heterogeneity is indeed present and treatment effects are subpopulation specific, the average treatment effect obtained by
the standard methods can lead to biased and incorrect conclusions. Hence the use of mixture models to represent the response distribution within the treatment group is compelling, and it is desirable to describe the nature of this subpopulation specific effect by estimating the corresponding parameters from the mixture distribution. 
	
	Denote the cumulative distribution functions associated with a response from the control group and the treatment group by $F$ and $G$, respectively. Shift alternatives of the form $G(u)= F(u-\delta)$, for a specified $\delta$, are frequently used. 
	In this paper we assume, without loss of generality, that $\delta > 0$ and we use a mixture model for the responses from the treatment group of the form
	\begin{equation}
		G(u)= (1-\theta) F(u) + \theta F(u-\delta), \label{mod}
	\end{equation}		
where $F \in \mathcal{C}, \theta \in (0,1]$, and $\mathcal{C}$ represents the space of absolutely continuous cumulative distribution functions (cdfs). Thus, the parameter space is $\Theta \equiv (\mathcal{C},(0,1],\mathcal{R^+})$. In this context, the treatment effect is represented by the pair $(\theta, \delta)$ and the average treatment effect for the whole population is $\Delta = \theta \delta$. (Note that when $\theta = 1$ the model simplifies to a simple mean shift).
		
		In this paper we assume the availability of an iid sample $X_1,...,X_m$ from $F$ and an iid sample $Y_1,...,Y_n$ from $G$ where the $X$'s and $Y$'s are independent for a total sample size of $N = m + n$. Denoting the mean of $F$ and $G$ by $\mu_{X}$ and $\mu_{Y}$, respectively, we have $\mu_Y = \mu_X + \Delta$ and therefore a (modified) method of moment estimator for $\Delta$ is $\hat{\Delta} = (\bar{Y} - \bar{X})_{+}$ where $t_+ = t$ if $t > 0$ and $0$ otherwise. The $+$ operator restricts $\hat{\Delta}$ to remain in the parameter space. Jeske and Yao$^7$ further proposed method of moment estimators for $(\theta,\delta)$ in \eqref{mod} of the form
	\begin{equation}
	\hat{\theta} = \left\{ 1 + \frac{(S^2_Y - S^2_X)_+}{(\bar{Y} - \bar{X})^2_{+} + \epsilon_N} \right\}^{-1} \label{theta.hat}
	\end{equation}
	\begin{equation}
	\hat{\delta} = (\bar{Y} - \bar{X})_{+} \cdot \left \{ 1 + \frac{(S^2_Y - S^2_X)_+}{(\bar{Y} - \bar{X})^2_{+} + \epsilon_N} \right \}, \label{delta.hat}
	\end{equation}
where $(\bar{X},S_{X}^2)$ are the sample mean and variance of the control group observations, $(\bar{Y},S_{Y}^2)$ are the same for the treatment group observations, and $\epsilon_N=o_p(1)$ is a small positive number that bounds the denominators away from zero. 
Based on our empirical studies, 
the choice of $\epsilon_N = S_X\log(N^2)/N$ works well and is used for all simulations in this paper. One nice property of the method of moment estimators \eqref{theta.hat} and \eqref{delta.hat} is that they provide consistent estimators under very weak conditions and do not require any parametric assumptions about the distribution $F(u)$.

	
The main objective of this paper is to investigate asymptotic properties of \eqref{theta.hat} and \eqref{delta.hat} and discuss how to construct confidence intervals for the treatment effect $(\theta,\delta)$ when the treatment population is not homogeneous.	
Our numerical studies demonstrate that the confidence intervals built based on our asymptotic results perform comparably to and even better in many cases than the more computationally intensive bootstrap intervals and their Bias-Corrected versions.$^{8,9}$

The rest of this paper is organized as follows. In section 2 we discuss the identifiability of the model (\ref{mod}) and derive the first four moments of $Y$ in terms of $(F,\theta,\delta)$, which will be used in section 3 where we discuss consistency and asymptotic normality of the method of moment estimators. We utilize these properties to propose large sample confidence intervals in section 4, where we further investigate how large the sample sizes $(m, n)$ need to be in order for the intervals to be accurate enough to provide satisfactory confidence intervals and compare the performance with Bias-Corrected Bootstrap Intervals from a simulation study. In section 5 we summarize the results and in section 6 we discuss future work.
	
\section{Preliminaries}
	\subsection{Identifiability}
	Note that the model (\ref{mod}) is generally not identifiable without imposing any shape or parametric assumption about $F(u)$ since $F$ itself could be a mixture distribution. However, for our semi-supervised problem, the control data is from $F(u)$ and hence $F$ is identifiable. Based on Jeske and Yao$^7$ (Proposition 4.1), the parameter $\theta$ and $\delta$ in (\ref{mod}) can be uniquely identified from the population moments of $F(u)$ and $G(u)$. Therefore, the model (\ref{mod}) is identifiable in our setting even without any parametric or shape assumption about $F(u)$.


	\subsection{Moments}
	In this section we derive formulas for some moments of $G$ in terms of $(F,\theta,\delta)$. We utilize these results to derive the variance of (\ref{theta.hat}) and (\ref{delta.hat}) in Section 3. Let $\mu_X = E\left[X\right], \sigma^2_X = E\left[(X-\mu_X)^2\right], \mu_{3c_X} = E[(X-\mu_X)^3],$ and $\mu_{4c_X} = E[(X-\mu_X)^4]$. Similarly, let $\mu_Y = E[Y], \sigma^2_Y = E[(Y-\mu_Y)^2], \mu_{3c_Y} = E[(Y-\mu_Y)^3],$ and $\mu_{4c_Y} = E[(Y-\mu_Y)^4]$.

\begin{prop}
	For  $(F,\theta,\delta) \in \Theta \equiv (\mathcal{C},(0,1],\mathcal{R^+})$ with $X \sim F$ having finite fourth moment, the moments of $Y \sim G$ can be found in terms of $(F,\theta,\delta)$ and are as follows
\begin{align}
	\mu_Y &= \mu_X + \theta \delta \label{m1} \\
	\sigma^2_Y &= \sigma^2_X + \theta (1-\theta) \delta^2 \label{m2} \\
	\mu_{3c_Y} &= \mu_{3c_X} + \theta (1-\theta) \delta^3 \left[1 - 2\theta \right] \label{m3} \\
	\mu_{4c_Y} &= \mu_{4c_X} + \theta (1-\theta) \delta^4 \left[(1-\theta)(1-3\theta) + \theta + 6\sigma_X^2/\delta^2 \right]. \label{m4}
\end{align}
\end{prop}
	Equations ($\ref{m1}$) - ($\ref{m4}$) are proved in Appendix A. Notice that each moment of $Y \sim G$ can be written in terms of the corresponding moment of $X \sim F$ plus a term that depends on $(\theta,\delta)$. The even central moments of $Y$ - (\ref{m2}) and (\ref{m4}) - can be minimized by letting $\delta$ become arbitrarily small or letting $\theta$ approach either $0$ or $1$ as the additional terms are non-negative. Such cases characterize a scenario where the treatment group's response distribution approaches (a potentially shifted version of) the control group's response distribution. The difference $\mu_{3c_{Y}} - \mu_{3c_{X}}$ may be positive or negative, and  will be $0$ when $\theta \in \{.5,1\}$ or as $\theta$ approaches $0$.

\section{Consistency and Asymptotic Normality}
	In this section we show that $\hat{\theta}$ and $\hat{\delta}$ in \eqref{theta.hat} and \eqref{delta.hat} are consistent and asymptotically normal estimators of $\theta$ and $\delta$ respectively.
	
\begin{prop}
For any $(F,\theta,\delta) \in \Theta \equiv \left(\mathcal{C},(0,1],\mathcal{R^+}\right)$ where $F$ has a finite second moment, $\hat{\theta} \overset{p}{\rightarrow} \theta$ and $\hat{\delta} \overset{p}{\rightarrow} \delta$.
\end{prop}
Specifically, in the case of $m = n$ we have the following result.
\begin{prop}
\label{prop:norm}
For any $(F,\theta,\delta) \in \left(\mathcal{C},(0,1),\mathcal{R^+}\right)$ where $F$ has a finite fourth moment
\begin{align*}
	\sqrt{n}(\hat{\theta} - \theta) \rightarrow N(0,\sigma^2_{\theta}) \\
	\sqrt{n}(\hat{\delta} - \delta) \rightarrow N(0,\sigma^2_{\delta}),
\end{align*}
where 
\begin{align}
	\nonumber \sigma^2_{\theta} &= \left(1 + \frac{\sigma^2_Y - \sigma^2_X}{(\mu_Y - \mu_X)^2}\right)^{-4}
	\Bigg \{
	\frac{4(\sigma^2_Y - \sigma^2_X)^2}{(\mu_Y - \mu_X)^6}\left(\sigma^2_X + \sigma^2_Y \right) 
	- \ \frac{4(\sigma^2_Y - \sigma^2_X)}{(\mu_Y - \mu_X)^5}\left(\mu_{3c_X} + \mu_{3c_Y}\right) \\
	& + \ \frac{\left(\mu_{4c_X} - \sigma^4_X \right) + \left( \mu_{4c_Y} - \sigma^4_Y \right)}{(\mu_Y - \mu_X)^4}
	\Bigg \} \label{sig2t} \\
	\nonumber \sigma^2_{\delta} &= \left(1 - \frac{\sigma^2_Y - \sigma^2_X}{(\mu_Y - \mu_X)^2} \right)^2 \left(\sigma^2_X + \sigma^2_Y \right)
	+ \ 2\left(1 - \frac{\sigma^2_Y - \sigma^2_X}{(\mu_Y - \mu_X)^2}\right) \left( \frac{\mu_{3c_X} + \mu_{3c_Y}}{\mu_Y - \mu_X} \right) \\
	& + \ \frac{\left(\mu_{4c_X} - \sigma^4_X\right) + \left(\mu_{4c_Y} - \sigma^4_Y\right)}{(\mu_Y - \mu_X)^2}. \label{sig2d}
\end{align}
\end{prop}
Propositions 3.1 and 3.2 are proved in Appendix B.

    \begin{figure*}[h]
        \centering
        \begin{subfigure}[b]{0.475\textwidth}
            \centering
            \includegraphics[width=\textwidth]{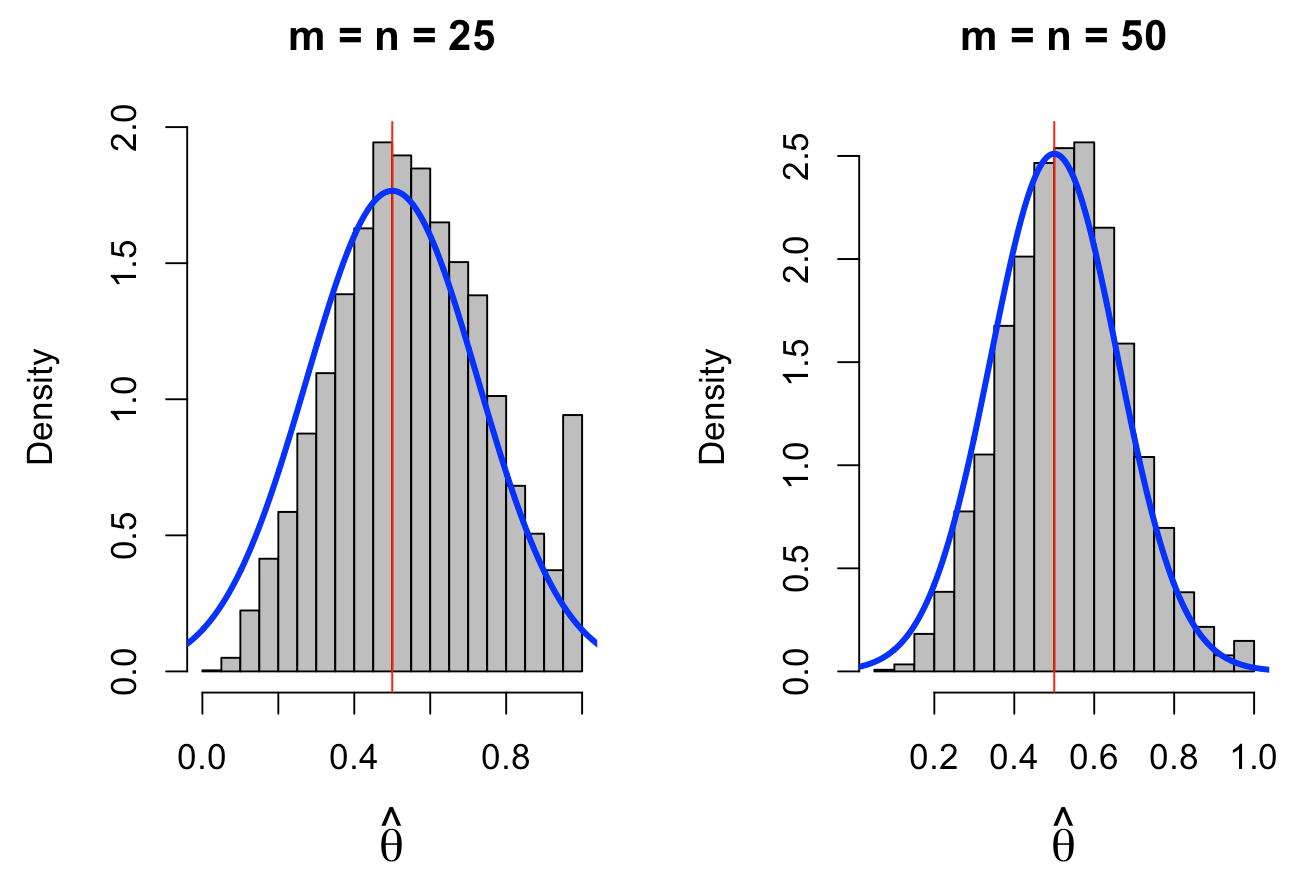}
            \label{fig:dist1}
        \end{subfigure}
        \hfill
        \begin{subfigure}[b]{0.475\textwidth}  
            \centering 
            \includegraphics[width=\textwidth]{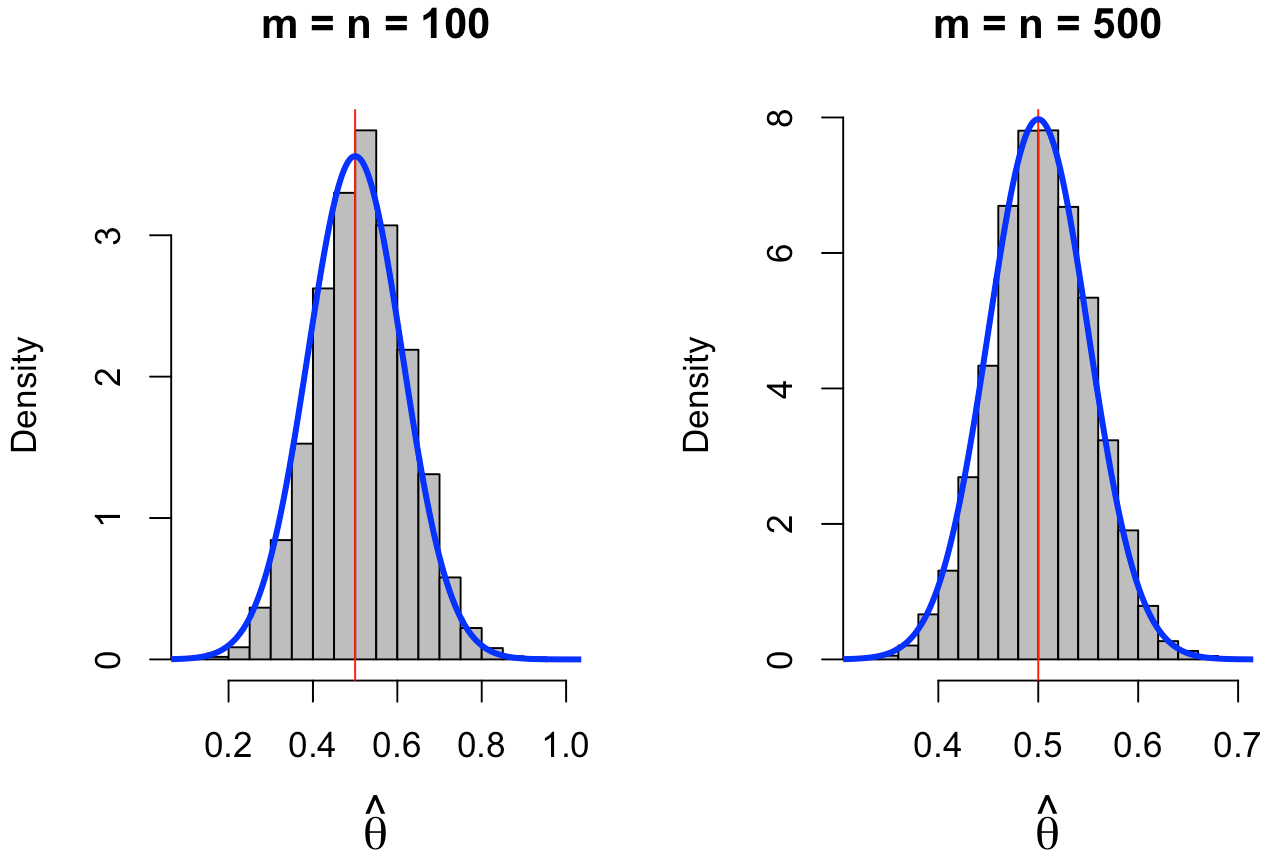}
            \label{fig:dist2}
        \end{subfigure}
        \vskip\baselineskip
        \begin{subfigure}[b]{0.475\textwidth}   
            \centering 
            \includegraphics[width=\textwidth]{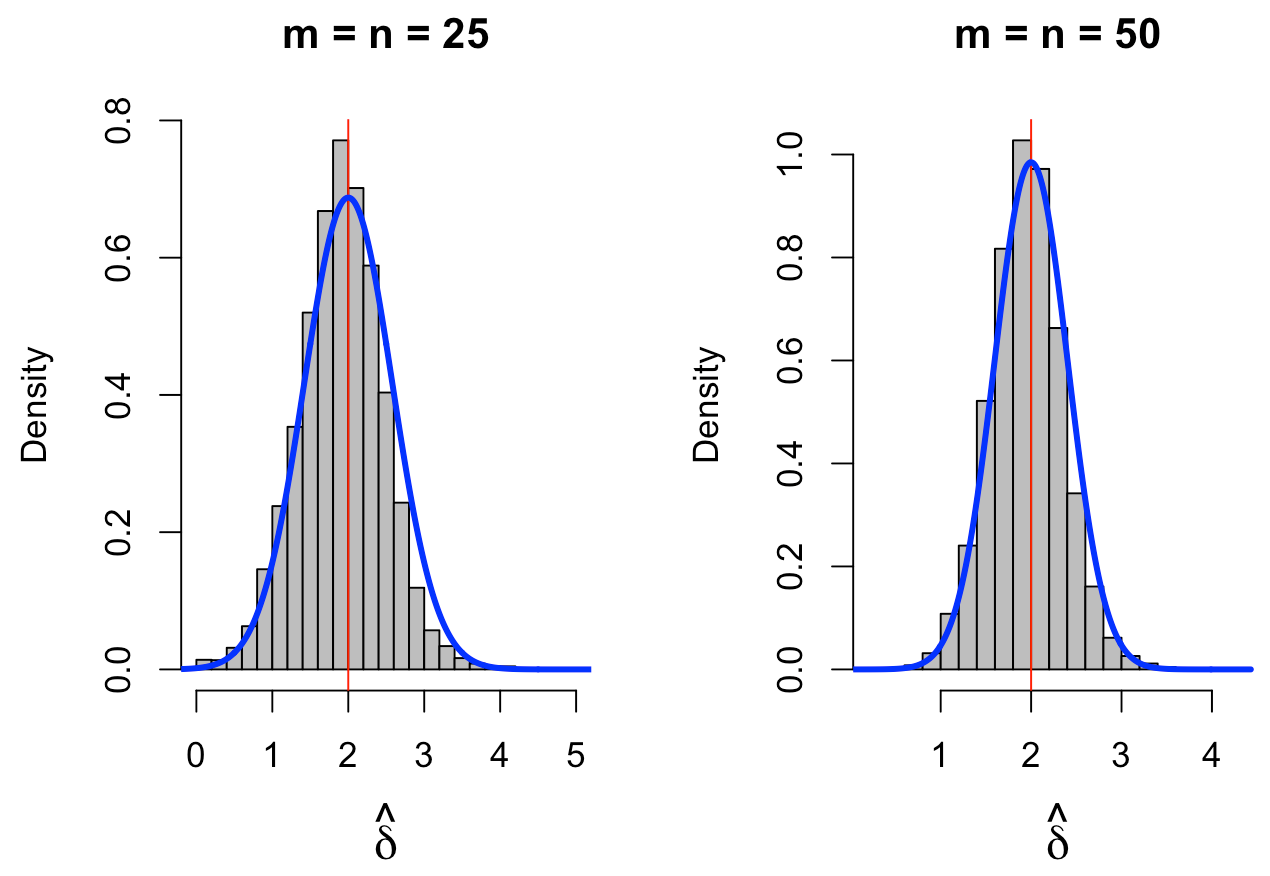}
            \label{fig:dist3}
        \end{subfigure}
        \quad
        \begin{subfigure}[b]{0.475\textwidth}   
            \centering 
            \includegraphics[width=\textwidth]{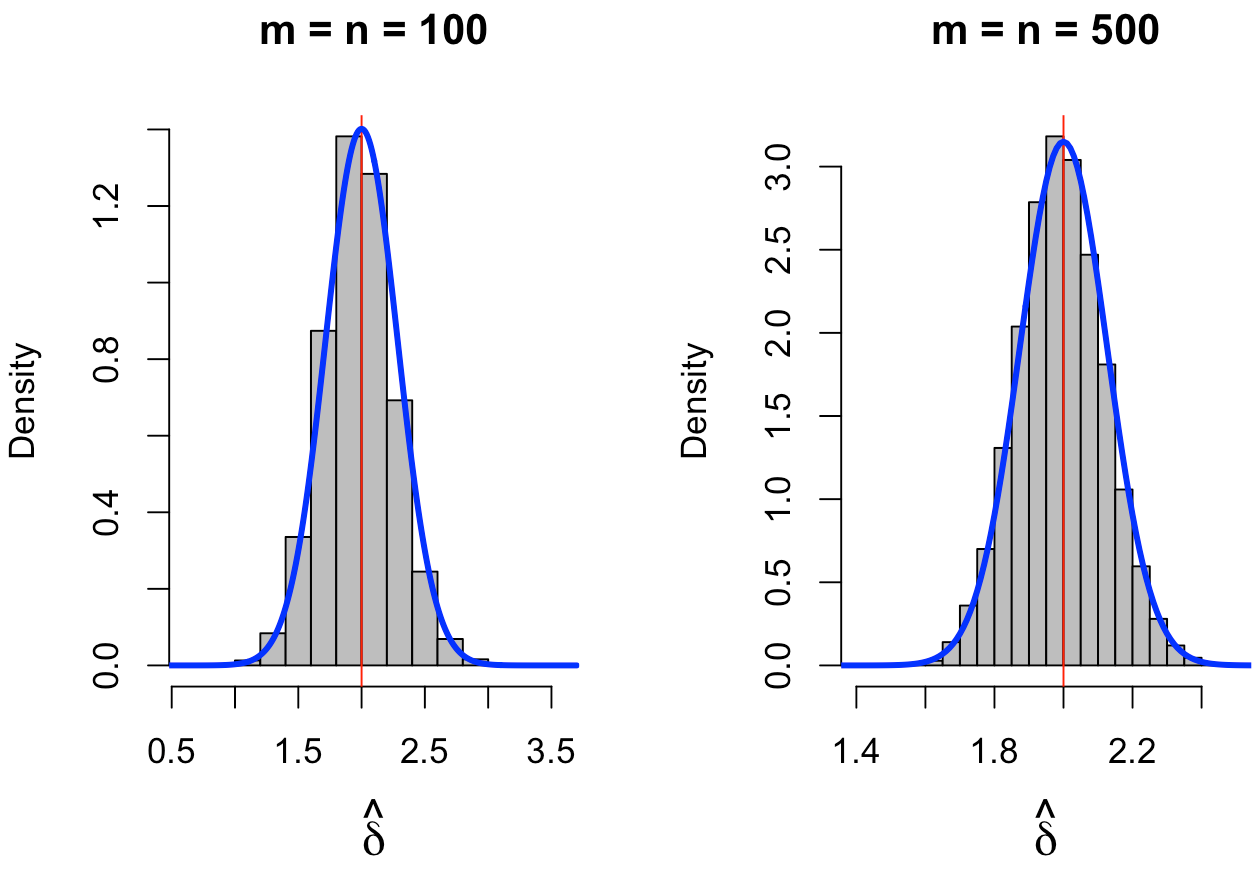}
            \label{fig:dist4}
        \end{subfigure}
        \caption[ Approximate Normality of Estimator Distributions]
        {\small Illustration of asymptotic normality for distributions of $\hat{\theta}$ and $\hat{\delta}$ for $F \sim Laplace$, $\sigma_{X} = 1$, $\theta = .5$, $\delta = 2$ and sample sizes $m=n \in \{ 25, 50, 100, 500 \}$. Blue curve represents approximate distribution based on Proposition 3.2.}
        \label{fig:norm}
    \end{figure*}
    
    \text{Figure \ref{fig:norm}} illustrates  the convergence of $\hat{\theta}$ and $\hat{\delta}$ in distribution to normal. For the selected parameter settings we can see the lack of normality due to the bounding of $\hat{\theta} \leq 1$ in the top left plot of the figure, which subsides as the sample sizes increase. We also see the elimination of the slight positive bias in $\hat{\theta}$ and negative bias in $\hat{\delta}$ as the sample sizes increase.

    
\section{Confidence Intervals}
	In this section we compare the performance of various confidence intervals based on the method of moment estimators presenting performance statistics for two of them - asymptotic confidence intervals and bootstrap bias-corrected acceleration (BCa) intervals$^8$. The asymptotic confidence intervals rely on the asymptotic normality of $\hat{\theta}$ and $\hat{\delta}$ presented in Proposition \ref{prop:norm}. Since, for sufficiently large $(m,n)$ - here considering $m=n$ - we have $\hat{\theta} \ \dot\sim \ N(\theta,\sigma^2_{\theta}/n)$ and $\hat{\delta} \ \dot\sim \ N(\delta,\sigma^2_{\delta}/n)$, where $\dot\sim$ means ``is approximately distributed as", the proposed asymptotic 100(1-$\alpha$)\% CIs for $\theta$ and $\delta$, respectively, are
\begin{align}
	CI_{asy}(\theta) = \left(\hat{\theta} - z_{\alpha/2} \cdot \hat{\sigma}_{\theta}/\sqrt{n} , \ \hat{\theta} + z_{\alpha/2} \cdot \hat{\sigma}_{\theta}/\sqrt{n} \right) \\
	CI_{asy}(\delta) = \left(\hat{\delta} - z_{\alpha/2} \cdot \hat{\sigma}_{\delta}/\sqrt{n} , \ \hat{\delta} + z_{\alpha/2} \cdot \hat{\sigma}_{\delta}/\sqrt{n} \right),
\end{align}
where $z_{\alpha/2} = \Phi^{-1}(1-\alpha/2)$. The standard errors, $\hat{\sigma}_{\theta}$ and $\hat{\sigma}_{\delta}$, are found by plugging in the sample moments as estimates for the population moments found in the asymptotic variance formulas and making the same alterations as in the estimators. That is, $\left(\mu_{Y} - \mu_{X}\right)$ is estimated with $\left(\bar{Y} - \bar{X}\right)_{+} + \epsilon_{N}$ and $\left(\sigma^2_{Y} - \sigma^2_{X} \right)$ is estimated with $\left(S^2_{Y} - S^2_{X}\right)_{+}$. Finally, when necessary the boundaries of the asymptotic confidence interval are truncated at the edge of the parameter spaces. For the bootstrap BCa CIs in this setting, we implement the following
\begin{enumerate}
	\item Randomly sample from $X_{1},...,X_{m}$ and $Y_{1},...,Y_{n}$ independently with replacement B=1000 times.
	\item For each of these 1000 bootstrap samples, calculate $\hat{\theta}_{b}$ and $\hat{\delta}_{b}$ to obtain bootstrap sampling distributions.
	\item Calculate the bias ($z_{0}$) based on \eqref{z} and acceleration ($a$) correction terms for both $\hat{\theta}$ and $\hat{\delta}$ respectively based on \eqref{eq:defa}.
	\item Calculate the percentiles of the bootstrap distributions to use for the confidence interval based on \eqref{lbci} and \eqref{upci}.
\end{enumerate}
Typically$^8$ in step 3, for the generic parameter $\tau$ - that is, either $\theta$ or $\delta$ - we have that $z_{0}$ and $a$ are calculated as $z_{0} = \Phi^{-1} \left( \# \left\{ \hat{\tau_{b}} < \hat{\tau} \right\}/B \right)$ and 
\begin{equation}
    a = \frac{\sum_{i=1}^{n} \left(\bar{\tau}_{b} - \hat{\tau}_{(-i)} \right)^{3} }{ 6 \{ \sum_{i=1}^{n} \left(\bar{\tau}_{b} - \hat{\tau}_{(-i)} \right)^{2} \}^{3/2} },
    \label{eq:defa}
\end{equation}
where $\#$ is the counting operator and $\hat{\tau}_{(-i)}$ is the estimate with the ith observation removed. However, this formula for $z_{0}$ can fail for $\hat{\theta}$ or $\hat{\delta}$ because of the bounded nature of the parameter space, and thus the estimators. There is non-zero probability that $\hat{\delta} = 0$, in which case $z_{0} = -\infty$. We propose adjusting for the discrete nature of the bootstrap sampling distributions by taking
\begin{align}
	z_{0}  = \Phi^{-1} \left( \left\{ \#\left(\hat{\tau}_{b} < \hat{\tau} \right) + \frac{1}{2} \#\left( \hat{\tau}_{b} = \hat{\tau} \right) \right\}/B \right). \label{z}
\end{align}
Step 4 remains unchanged where we let
\begin{align}
	\alpha_{l} = \Phi \left( z_{0} + \frac{z_{0} - z_{\alpha/2}}{1 - a(z_{0} - z_{\alpha/2})} \right) \label{lbci}\\
	\alpha_{u} = \Phi \left( z_{0} + \frac{z_{0} + z_{\alpha/2}}{1 - a(z_{0} + z_{\alpha/2})} \right),
	\label{upci}
\end{align}
giving the BCa interval $[\hat{\tau}^{(\alpha_{l})}_{b}, \hat{\tau}^{(\alpha_{u})}_{b}]$. We also investigated centered bootstrap percentile confidence intervals$^9$ but the results are not presented due to poor performance.  In the simulation the shift $\delta$ was parameterized as $\delta=K\sigma_X$. Based on Jeske and Yao$^7$, the performance of the estimators and confidence intervals only depend on $F$, $\theta$, and $K$.  Some performance statistics of the asymptotic and BCa confidence intervals are displayed for the following combinations of the parameters $m=n \in \{ 25, 50, 100, 500 \}$, $F \in \{ Normal, Logistic, Laplace \}$, $\theta \in \{.5, .8 \}$, and $K \in \{1, 3\}$. Tables 1 and 2 present the results of intervals for $\theta$  and $\delta$, respectively. 


\begin{table}[!ht]
\centering
\begin{tabular}{ |p{1cm} p{.4cm} p{0.4cm} | p{0cm}  p{.80cm} p{.80cm} | p{0cm} p{.80cm} p{.80cm} | p{0cm}  p{.80cm} p{.80cm} | p{0cm} p{.80cm} p{.80cm}| }
\hline
	 \multicolumn{3}{|c|}{Parameters} & & \multicolumn{2}{c|}{ Asy Int} & & \multicolumn{2}{c|}{ BCa Int} & & \multicolumn{2}{c|}{ Asy Int} & & \multicolumn{2}{c|}{ BCa Int} \\
  \hline
$F$ & $\theta$ & $K$ &  & Cov. Prob & Avg. Len & & Cov. Prob & Avg. Len & & Cov. Prob & Avg. Len & & Cov. Prob & Avg. Len \\ 
\hline
 	\multicolumn{3}{|c|}{} & &  \multicolumn{5}{c|}{m = n = 25} & & \multicolumn{5}{c|}{m = n = 50} \\
  \hline
  Normal & 0.5 & 1 & & 0.95 & 0.87 & & 0.79 & 0.68 & & 0.97 & 0.85 & & 0.81 & 0.69 \\ 
  Normal & 0.5 & 3 & & 0.9 & 0.55 & & 0.95 & 0.55 & & 0.93 & 0.42 & & 0.94 & 0.42 \\ 
  Normal & 0.8 & 1 & & 0.95 & 0.76 & & 0.8 & 0.54 & & 0.96 & 0.71 & & 0.81 & 0.52 \\ 
  Normal & 0.8 & 3 & & 0.86 & 0.35 & & 0.91 & 0.43 & & 0.9 & 0.28 & & 0.95 & 0.32 \\ 
  Logistic & 0.5 & 1 & & 0.95 & 0.87 & & 0.73 & 0.66 & & 0.95 & 0.86 & & 0.8 & 0.68 \\ 
  Logistic & 0.5 & 3 & & 0.91 & 0.55 & & 0.93 & 0.55 & & 0.92 & 0.42 & & 0.95 & 0.42 \\ 
  Logistic & 0.8 & 1 & & 0.94 & 0.78 & & 0.79 & 0.54 & & 0.94 & 0.75 & & 0.8 & 0.52 \\ 
  Logistic & 0.8 & 3 & & 0.88 & 0.37 & & 0.92 & 0.42 & & 0.9 & 0.29 & & 0.93 & 0.32 \\ 
  Laplace & 0.5 & 1 & & 0.94 & 0.87 & & 0.72 & 0.64 & & 0.93 & 0.86 & & 0.79 & 0.67 \\ 
  Laplace & 0.5 & 3 & & 0.89 & 0.56 & & 0.94 & 0.56 & & 0.93 & 0.43 & & 0.94 & 0.42 \\ 
  Laplace & 0.8 & 1 & & 0.94 & 0.8 & & 0.8 & 0.54 & & 0.93 & 0.79 & & 0.8 & 0.51 \\ 
  Laplace & 0.8 & 3 & & 0.89 & 0.37 & & 0.91 & 0.42 & & 0.93 & 0.3 & & 0.94 & 0.31 \\ 
  \hline
	\multicolumn{3}{|c|}{} & &  \multicolumn{5}{c|}{m = n = 100} & & \multicolumn{5}{c|}{m = n = 500} \\
\hline
  Normal & 0.5 & 1 & & 0.96 & 0.81 & & 0.83 & 0.67 & & 0.96 & 0.48 & & 0.94 & 0.5 \\ 
  Normal & 0.5 & 3 & & 0.93 & 0.3 & & 0.95 & 0.31 & & 0.94 & 0.14 & & 0.96 & 0.14 \\ 
  Normal & 0.8 & 1 & & 0.97 & 0.6 & & 0.78 & 0.47 & & 0.97 & 0.34 & & 0.93 & 0.33 \\ 
  Normal & 0.8 & 3 & & 0.94 & 0.21 & & 0.95 & 0.23 & & 0.95 & 0.1 & & 0.95 & 0.1 \\ 
  Logistic & 0.5 & 1 & & 0.95 & 0.83 & & 0.84 & 0.67 & & 0.97 & 0.55 & & 0.94 & 0.5 \\ 
  Logistic & 0.5 & 3 & & 0.93 & 0.31 & & 0.94 & 0.3 & & 0.95 & 0.14 & & 0.94 & 0.14 \\ 
  Logistic & 0.8 & 1 & & 0.94 & 0.66 & & 0.8 & 0.46 & & 0.95 & 0.39 & & 0.91 & 0.32 \\ 
  Logistic & 0.8 & 3 & & 0.93 & 0.22 & & 0.95 & 0.22 & & 0.94 & 0.1 & & 0.95 & 0.1 \\ 
  Laplace & 0.5 & 1 & & 0.94 & 0.84 & & 0.84 & 0.67 & & 0.96 & 0.63 & & 0.94 & 0.49 \\ 
  Laplace & 0.5 & 3 & & 0.95 & 0.32 & & 0.96 & 0.3 & & 0.94 & 0.15 & & 0.96 & 0.14 \\ 
  Laplace & 0.8 & 1 & & 0.93 & 0.73 & & 0.81 & 0.47 & & 0.95 & 0.45 & & 0.92 & 0.33 \\ 
  Laplace & 0.8 & 3 & & 0.93 & 0.23 & & 0.94 & 0.22 & & 0.96 & 0.11 & & 0.95 & 0.1 \\
  \hline
\end{tabular}
\label{tab:CIth}
\caption{Coverage Probabilities and Average Lengths of Asymptotic and BCa 95\% Confidence Intervals for $\theta$ based upon 1000 data sets per setting.}
\end{table}

Table 1 shows that the coverage probability of the 95\% asymptotic interval for $\theta$ is well calibrated except for the case of very small sample sizes $m=n=25$. However, the BCa intervals have far too low coverage probabilities when $K$ is small even for moderate sample size, say $m=n=100$, but well-calibrated coverage probabilities for large $K$. As the sample sizes increase both confidence intervals have coverage probabilities converging toward .95 but the asymptotic interval appears to do so more quickly. The lengths of intervals vary widely as a function of the parameters. Under many settings, the expected length of the confidence interval for $\theta$ may be too large to provide clinically meaningful information. Generally speaking, smaller $\theta, K, m=n$ all result in increased interval lengths for both methods, with $F$ having a minimal effect. When the component distributions are not well separated, increasing the sample size seems to have a slow effect on reducing the interval lengths. 

\begin{table}[!ht]
\centering
\begin{tabular}{ |p{1cm} p{.4cm} p{.3cm} | p{0cm}  p{.80cm} p{.80cm} | p{0cm} p{.80cm} p{.80cm} | p{0cm}  p{.80cm} p{.80cm} | p{0cm} p{.80cm} p{.80cm}| }
\hline
	 \multicolumn{3}{|c|}{Parameters} & & \multicolumn{2}{c|}{ Asy Int} & & \multicolumn{2}{c|}{ BCa Int} & & \multicolumn{2}{c|}{ Asy Int} & & \multicolumn{2}{c|}{ BCa Int} \\
  \hline
$F$ & $\theta$ & $K$ &  & Cov. Prob & Avg. Len & & Cov. Prob & Avg. Len & & Cov. Prob & Avg. Len & & Cov. Prob & Avg. Len \\ 
  \hline
  	\multicolumn{3}{|c|}{} & &  \multicolumn{5}{c|}{m = n = 25} & & \multicolumn{5}{c|}{m = n = 50} \\
\hline
Normal & 0.5 & 1 & & 0.98 & 2.84 & & 0.87 & 1.77 & & 0.99 & 2.23 & & 0.86 & 1.54 \\ 
  Normal & 0.5 & 3 & & 0.93 & 2.01 & & 0.92 & 2.24 & & 0.95 & 1.37 & & 0.93 & 1.48 \\ 
  Normal & 0.8 & 1 & & 0.94 & 1.94 & & 0.93 & 1.5 & & 0.98 & 1.48 & & 0.94 & 1.16 \\ 
  Normal & 0.8 & 3 & & 0.94 & 1.3 & & 0.92 & 1.36 & & 0.93 & 0.92 & & 0.93 & 0.94 \\ 
  Logistic & 0.5 & 1 & & 0.98 & 3.18 & & 0.88 & 1.76 & & 0.99 & 2.49 & & 0.84 & 1.53 \\ 
  Logistic & 0.5 & 3 & & 0.94 & 2.07 & & 0.92 & 2.24 & & 0.95 & 1.47 & & 0.95 & 1.49 \\ 
  Logistic & 0.8 & 1 & & 0.95 & 2.07 & & 0.93 & 1.51 & & 0.96 & 1.63 & & 0.92 & 1.17 \\ 
  Logistic & 0.8 & 3 & & 0.94 & 1.35 & & 0.9 & 1.33 & & 0.94 & 0.96 & & 0.94 & 0.95 \\ 
  Laplace & 0.5 & 1 & & 0.98 & 3.63 & & 0.88 & 1.79 & & 0.99 & 2.96 & & 0.84 & 1.54 \\ 
  Laplace & 0.5 & 3 & & 0.93 & 2.22 & & 0.92 & 2.27 & & 0.94 & 1.55 & & 0.94 & 1.48 \\ 
  Laplace & 0.8 & 1 & & 0.95 & 2.22 & & 0.93 & 1.52 & & 0.97 & 1.85 & & 0.93 & 1.16 \\ 
  Laplace & 0.8 & 3 & & 0.93 & 1.39 & & 0.92 & 1.35 & & 0.95 & 1.03 & & 0.93 & 0.94 \\ 
\hline
	\multicolumn{3}{|c|}{} & &  \multicolumn{5}{c|}{m = n = 100} & & \multicolumn{5}{c|}{m = n = 500} \\
\hline
  Normal & 0.5 & 1 & & 0.99 & 1.66 & & 0.83 & 1.28 & & 0.96 & 0.78 & & 0.94 & 0.75 \\ 
  Normal & 0.5 & 3 & & 0.95 & 0.94 & & 0.94 & 1 & & 0.96 & 0.42 & & 0.95 & 0.43 \\ 
  Normal & 0.8 & 1 & & 0.97 & 1.09 & & 0.92 & 0.88 & & 0.98 & 0.5 & & 0.92 & 0.46 \\ 
  Normal & 0.8 & 3 & & 0.95 & 0.66 & & 0.94 & 0.67 & & 0.94 & 0.29 & & 0.94 & 0.3 \\ 
  Logistic & 0.5 & 1 & & 0.99 & 1.88 & & 0.83 & 1.29 & & 0.97 & 0.94 & & 0.94 & 0.75 \\ 
  Logistic & 0.5 & 3 & & 0.95 & 1.02 & & 0.94 & 0.99 & & 0.94 & 0.45 & & 0.95 & 0.43 \\ 
  Logistic & 0.8 & 1 & & 0.96 & 1.27 & & 0.9 & 0.88 & & 0.96 & 0.6 & & 0.91 & 0.46 \\ 
  Logistic & 0.8 & 3 & & 0.94 & 0.69 & & 0.94 & 0.66 & & 0.96 & 0.31 & & 0.95 & 0.3 \\ 
  Laplace & 0.5 & 1 & & 0.99 & 2.15 & & 0.84 & 1.29 & & 0.98 & 1.14 & & 0.94 & 0.76 \\ 
  Laplace & 0.5 & 3 & & 0.94 & 1.1 & & 0.94 & 1.01 & & 0.95 & 0.51 & & 0.96 & 0.43 \\ 
  Laplace & 0.8 & 1 & & 0.97 & 1.46 & & 0.93 & 0.88 & & 0.97 & 0.73 & & 0.9 & 0.46 \\ 
  Laplace & 0.8 & 3 & & 0.95 & 0.76 & & 0.93 & 0.66 & & 0.96 & 0.34 & & 0.97 & 0.3 \\ 
   \hline
\end{tabular}
\label{tab:CIdel}
\caption{Coverage Probabilities and Average Lengths of Asymptotic and BCa 95\% Confidence Intervals for $\delta$ based upon 1000 data sets per setting.}
\end{table}

Table 2 shows that the coverage probability for the 95\% asymptotic interval for $\delta$ tends to be conservative when the component distributions are not well separated and are fairly well-calibrated otherwise even for small sample sizes. Contrarily, the BCa confidence intervals tend to have coverage probabilities that are too low and this is most notable when the components are not well separated. As the sample sizes increase, both methods have coverage probabilities that converge to .95 rather slowly when $K$ is small. When $\theta, K, m=n$ are small and $F$ is heavy-tailed, the interval lengths for $\delta$ tend to be quite large. The BCa interval lengths are more robust to heavy-tailed $F$ than the asymptotic intervals. For both intervals, increasing the sample sizes has a more notable impact on decreasing interval length for $\delta$ than for $\theta$. Sample sizes may need to be large for interval lengths to be clinically informative. 

\section{Summary}
In applications where a subset of the population will respond to the treatment, the mixture of the control response distribution and a shift of that distribution is useful for characterizing both the proportion of responders in the population as well as the effect of the treatment on the responders. The asymptotic formulas derived here based upon method of moment estimators prove useful for constructing confidence intervals for both $\theta$ and $\delta$ as they are easier to compute and their performance is shown to be comparable to or better than both centered bootstrap percentile confidence intervals as well as the bias-corrected accelerated bootstrap confidence intervals. These intervals will be narrow enough to provide meaningful insight for sufficiently large trials such as in knoll et al.$^{10}$

\section{Future Work}
 In this paper, we consider method of moment estimators for \eqref{mod} since they are easy to compute and free of parametric assumption of $F$. It will be interesting to investigate whether we could extend the semiparametric efficient estimator of Ma et al.$^{11}$ to get a more efficient estimator of $(\theta, \delta)$. Since the treatment effect is characterized by the pair $(\theta, \delta)$ and the individual confidence intervals for $\theta$ and $\delta$ tend to be wide, we also wish to explore confidence regions for $(\theta, \delta)$ which may provide a more informative characterization of the treatment effect. 

\newpage

\section*{Appendix A: Proof of Proposition 2.1}

To derive equations (\ref{m1}) - (\ref{m4}), we use the following relationship from model (\ref{mod})
\begin{align*}
	Y \overset{d}{=} (1-Z)X + Z(X+\delta)
\end{align*}
where $Z \sim bernoulli(\theta)$ independent of $X \sim F$. This relationship holds because $Y \sim G$. Thus,
\begin{align}
	\nonumber \mu_Y \equiv E[Y] &= E\left[(1-Z)X + Z(X + \delta)\right] \\
	&= \mu_X + \theta \delta \tag{\ref{m1}}
\end{align}
To calculate (\ref{m2}), we first attain $E[Y^2]$ in terms of $(F,\theta,\delta)$. Letting $a = (1-Z)X$ and $b = Z(X+\delta)$
\begin{align*}
	E[Y^2] = E \left[ \left(a+b\right)^2 \right] &= E\left[ \left(a^2 + ab + b^2\right) \right] \\
	&= E \left[ a^2 + b^2 \right] \\
	&= \left(1-\theta\right)E[X^2] + \theta E\left[ E[X^2] + 2\mu_X\delta + \delta^2 \right] \\
	&= E[X^2] + 2\mu_X \theta \delta + \theta \delta^2.
\end{align*}
Notice that the terms for which both $a$ and $b$ have a non-zero exponents - $k_a$ and $k_b$ - are 0 because $(1-Z)^{k_a}Z^{k_b} = 0$ with probability $1$ whenever $k_a > 0$ and $k_b > 0$. Then, we have that
\begin{align*}
	\sigma^2_Y \equiv E[(Y-\mu_Y)^2] &= E[Y^2] - E[Y]^2 \\
	&= E[X^2] + 2 \theta \delta \mu_X + \theta \delta^2 - (\mu^2_X + 2 \theta \delta \mu_X + \theta^2 \delta^2) \\
	&= (E[X^2]  - \mu^2_X) + (\theta - \theta^2) \delta^2 \\
	&= \sigma^2_X + \theta (1-\theta) \delta^2 \tag{\ref{m2}}.
\end{align*}
To calculate (\ref{m3}), we first attain $E[Y^3]$ in terms of $(F,\theta,\delta)$.
\begin{align*}
	E[Y^3] &= E[(a+b)^3] \\
	&= E[a^3 + 3a^2b + 3ab^2 + b^3] \\
	&= E[a^3 + b^3] \\
	&= E[X^3 ]+ 3 \theta \delta E[X^2] + 3 \theta \delta^2 \mu_X + \theta \delta^3,
\end{align*}
again noting that $a^{k_a} b^{k_b} = 0$ if $k_a > 0$ and $k_b > 0$. Then, we have that
\begin{align*}
	\mu_{3cy} \equiv E[(Y-\mu_Y)^3] &= E[Y^3] - 3E[Y^2] \mu_Y + 2\mu^3_Y \\
	&= (E[X^3] - 3E[X^2]\mu_X + 2\mu^3_X) + \theta \delta^3 - 3\theta^2 \delta^3 + 2 \theta^3 \delta^3 \\
	&= \mu_{3cx} + \theta (1-\theta) \delta^3 [1 - 2\theta]. \tag{\ref{m3}}
\end{align*}
To calculate (\ref{m4}), we first attain $E[Y^4]$ in terms of $(F,\theta,\delta)$.
\begin{align*}
	E[Y^4] = E[(a+b)^4] &= E[a^4 + b^4] \\
	&= E[X^4] + 4\theta \delta E[X^3] + 6 \theta \delta^2 E[X^2] + 4 \theta \delta^3 \mu_X + \theta \delta^4,
\end{align*}
again noting that $a^{k_a} b^{k_b} = 0$ if $k_a > 0$ and $k_b > 0$. Then, we have that
\begin{align*}
	\mu_{4cy} \equiv E[(Y-\mu_Y)^4] &= E[Y^4] - 4\mu_Y E[Y^3] + 6\mu^2_YE[Y^2] - 3\mu^4_Y \\
	&= \left(E[X^4] - 4\mu_X E[X^3] + 6 \mu^2_X E[X^2] - 3\mu^4_X \right) \\
	&+ 6 \theta \delta^2 E[X^2] + \theta \delta^4 - 6 \theta \delta^2 \mu^2_X - 6 \theta^2 \delta^2 E[X^2] - 4 \theta^2 \delta^4 + 6 \theta^2 \delta^2 \mu^2_X + 6 \theta^3 \delta^4 - 3 \theta^4 \delta^4 \\
	&= \mu_{4cx} + \theta \delta^4 \left [ (1 - 4\theta + 6 \theta^2 - 3 \theta^3 ) + 6 (1-\theta) (\sigma^2_X/\delta^2) \right ] \\
	&= \mu_{4cx} + \theta \delta^4 \left [ \left( (1- 3\theta)(1-\theta)^2 + \theta (1-\theta) \right) + 6(1-\theta) (\sigma^2_X/\delta^2) \right] \\
	&= \mu_{4cx} + \theta (1-\theta) \delta^4 \left [ (1- \theta)(1-3\theta) + \theta + 6/K^2 \right], \tag{\ref{m4}}
\end{align*}
where $K=\delta/\sigma_X.$

\newpage

\section*{Appendix B: Proof of Proposition 3.1 and Proposition 3.2}

Proof of Proposition 3.1

Here we show the consistency of both $\hat{\theta}$ and $\hat{\delta}$ in estimating $\theta$ and $\delta$ respectively. First consider $f \left(\bar{X},\bar{Y},S^2_Y,S^2_X \right)$, an approximation of $\hat{\theta}$
\begin{align*}
\hat{\theta} = \left\{ 1 + \frac{(S^2_Y - S^2_X)_+}{(\bar{Y} - \bar{X})^2_{+} + \epsilon_N} \right\}^{-1} \approx \ \left\{ 1 + \frac{(S^2_Y - S^2_X)}{(\bar{Y} - \bar{X})^2} \right\}^{-1} \equiv f \left(\bar{X},\bar{Y},S^2_Y,S^2_X \right).
\end{align*}
If the sample sizes increase in such a way that both $m \rightarrow \infty$ and $n \rightarrow \infty$, then clearly $f \left(\bar{X},\bar{Y},S^2_Y,S^2_X \right) \overset{p}{\rightarrow} \theta$ since $\left(\bar{Y} - \bar{X} \right)^2 \overset{p}{\rightarrow} \left( \mu_{Y} - \mu_{X} \right)^2$, $\left(S^2_{Y} - S^2_{X} \right) \overset{p}{\rightarrow} \left(\sigma^2_{Y} - \sigma^2_{X} \right)^2$, and $\theta = \left\{ 1 + \frac{(\sigma^2_Y - \sigma^2_X)}{(\mu_{Y} - \mu_{X})^2} \right\}^{-1}$. Thus it suffices to show that $(S^2_Y - S^2_X)_+ \overset{p}{\rightarrow} (\sigma^2_{Y} - \sigma^2_{X})$ and $\left( \left(\bar{Y} - \bar{X} \right)^2_{+} + \epsilon_N \right) \overset{p}{\rightarrow} (\mu_{Y} - \mu_{X})$.
Since $(S^2_Y - S^2_X) \overset{p}{\rightarrow} (\sigma^2_{Y} - \sigma^2_{X})$, this means that $\forall \ \epsilon > 0$ and $\forall \ \omega > 0$, $\exists \ \{m_{0},n_{0}\}$ such that $\forall \ m > m_{0}$ and $\forall \ n > n_{0}$
\begin{align*}
	&P\left( \vert \left(S^2_Y - S^2_X\right) - \left(\sigma^2_{Y} - \sigma^2_{X}\right) \vert > \epsilon \right) < \omega \\
	\Leftrightarrow \ & P\left( \left(S^2_Y - S^2_X\right) - \left(\sigma^2_{Y} - \sigma^2_{X}\right) < -\epsilon \right) + P\left( \left(S^2_Y - S^2_X\right) - \left(\sigma^2_{Y} - \sigma^2_{X}\right) > \epsilon \right) < \omega.
\end{align*}
Also,
\begin{align*}
	&P\left( \vert \left(S^2_Y - S^2_X\right)_{+} - \left(\sigma^2_{Y} - \sigma^2_{X}\right) \vert > \epsilon \right) < \omega \\
	\Leftrightarrow \ & P\left( \left(S^2_Y - S^2_X\right)_{+} - \left(\sigma^2_{Y} - \sigma^2_{X}\right) < -\epsilon \right) + P\left( \left(S^2_Y - S^2_X\right)_{+} - \left(\sigma^2_{Y} - \sigma^2_{X}\right) > \epsilon \right) < \omega.
\end{align*}
So since
\begin{align*}
	P\left( \left(S^2_Y - S^2_X\right)_{+} - \left(\sigma^2_{Y} - \sigma^2_{X}\right) < -\epsilon \right) < P\left( \left(S^2_{Y} - S^2_{X} \right) - \left(\sigma^2_{Y} - \sigma^2_{X}\right) < -\epsilon \right),
\end{align*}
we have that
\begin{align*}
	(S^2_Y - S^2_X) \overset{p}{\rightarrow} (\sigma^2_{Y} - \sigma^2_{X}) \implies (S^2_Y - S^2_X)_+ \overset{p}{\rightarrow} (\sigma^2_{Y} - \sigma^2_{X}).
\end{align*}
An analogous argument shows that $\left(\bar{Y} - \bar{X} \right)^2_{+} \overset{p}{\rightarrow} (\mu_{Y} - \mu_{X})^2$. We have that $\left( \left(\bar{Y} - \bar{X} \right)^2_{+} + \epsilon_N \right) \overset{p}{\rightarrow} (\mu_{Y} - \mu_{X})^2$ and thus finally, $\hat{\theta} \overset{p}{\rightarrow} \theta$. The consistency of $\hat{\delta}$ immediately follows because $\hat{\delta} = \left(\bar{Y} - \bar{X}\right)_{+}/\hat{\theta}$ and $\delta = (\mu_{Y} - \mu_{X})/\theta$.

\newpage

\noindent Proof of Proposition 3.2

Here we prove asymptotic normality and derive the asymptotic variance of $\hat{\theta}$ and $\hat{\delta}$. We begin by proving Proposition 3.2 for $\hat{\theta}$,
\begin{align*}
\hat{\theta} = \left\{ 1 + \frac{(S^2_Y - S^2_X)_+}{(\bar{Y} - \bar{X})^2_{+} + \epsilon_N} \right\}^{-1} \approx \ \left\{ 1 + \frac{(S^2_Y - S^2_X)}{(\bar{Y} - \bar{X})^2} \right\}^{-1} \equiv f \left(\bar{X},\bar{Y},S^2_Y,S^2_X \right).
\end{align*}
Using a first order taylor series expansion we have that
	\begin{align*}
		f \left(\bar{X},\bar{Y},S^2_X,S^2_Y \right)  \approx f \left(\mu_X,\mu_Y,\sigma^2_X,\sigma^2_Y \right) \ & + \ \frac{\partial f}{\partial \bar{X}} \bigg\rvert_{\bar{X} = \mu_X} \left(\bar{X} - \mu_X \right) \ + \ \frac{\partial f}{\partial \bar{Y}} \bigg\rvert_{\bar{Y} = \mu_Y} \left(\bar{Y} - \mu_Y \right) \\
			& + \frac{\partial f}{\partial S^2_X} \bigg\rvert_{S^2_{X} = \sigma^2_X} \left(S^2_X - \sigma^2_X \right) \ + \ \frac{\partial f}{\partial S^2_Y} \bigg\rvert_{S^2_Y = \sigma^2_Y} \left(S^2_Y - \sigma^2_Y \right)
	\end{align*}
	\begin{align}
		\nonumber = f \left(\mu_{X},\mu_{Y},\sigma^2_{Y},\sigma^2_{X} \right) \ + \ \left\{1 + \frac{(\sigma^2_Y - \sigma^2_X)}{(\mu_Y - \mu_X)^2} \right\}^{-2} &\bigg\{ \ \frac{2(\sigma^2_Y - \sigma^2_X)(\mu_X - \bar{X})}{(\mu_Y - \mu_X)^3} \ + \ \frac{2(\sigma^2_Y - \sigma^2_X)(\bar{Y} - \mu_Y)}{(\mu_Y - \mu_X)^3} \\
		 &+ \ \frac{S^2_X - \sigma^2_X}{(\mu_Y - \mu_X)^2} \ + \ \frac{\sigma^2_Y - S^2_Y}{(\mu_Y - \mu_X)} \bigg\} \label{f}
	\end{align}
	
Now since $\bar{X}, \bar{Y}, S^2_X, S^2_Y$ are all unbiased estimators of $\mu_X, \mu_Y, \sigma^2_X, \sigma^2_Y$ respectively, we have that \\$E[f(\mu_X,\mu_Y,\sigma^2_X,\sigma^2_Y)] \approx \theta$ with accuracy to the first order expansion. Furthermore, since $\bar{X},\bar{Y},S^2_X,S^2_Y$ each converge in distribution to normal by the central limit theorem, have that $f\left(\bar{X},\bar{Y},S^2_X,S^2_Y \right)$ also converges in distribution to normal.

Now we derive the variance of $\hat{\theta}$ by taking the variance of (\ref{f}). First, note that any covariance terms between $X$ and $Y$ are 0 because $X$ and $Y$ are independent. Also, we utilize the following variance$^{12}$ and covariance$^{13,14}$ results in computing the variance of (\ref{f})
\begin{align}
	Var \left(S^2 \right) = \frac{1}{n} \cdot \left( \mu_{4c} - \frac{n-3}{n-1} \sigma^4 \right)
\end{align}
\begin{equation}
	Cov \left(\bar{X},S^2 \right) = \frac{\mu_{3c}}{n}.
\end{equation}
Thus, we have that the first order taylor series approximate variance of $\hat{\theta}$ is
\begin{align}
	\nonumber Var \left(\hat{\theta} \right) \approx \left(1 + \frac{\sigma^2_Y - \sigma^2_X}{(\mu_Y - \mu_X)^2} \right)^{-4} &\Bigg \{ \frac{4 \left(\sigma^2_Y - \sigma^2_X\right)^2}{(\mu_Y - \mu_X)^6} \left(\frac{\sigma^2_X}{m} + \frac{\sigma^2_Y}{n} \right) - \ \frac{4(\sigma^2_Y - \sigma^2_X)^2}{(\mu_Y - \mu_X)^5} \left( \frac{\mu_{3cx}}{m} + \frac{\mu_{3cy}}{n} \right) \\
	 &+ \ \frac{1}{(\mu_Y - \mu_X)^4} \left( \frac{(\mu_{4cx} - \frac{m-3}{m-1} \sigma^4_X)}{m} + \frac{(\mu_{4cy} - \frac{n-3}{n-1} \sigma^4_Y)}{n} \right) \Bigg\}.
\end{align}
The special case of $m=n$ gives the asymptotic variance formula in (\ref{sig2t}) as desired.

Now to prove Proposition 3.2 for the case of $\hat{\delta}$, we have
\begin{align*}
\hat{\delta} = \frac{\hat{\Delta}}{\hat{\theta}} = (\bar{Y} - \bar{X})_{+} \Big\{1 + \frac{(S^2_Y - S^2_X)_{+}}{(\bar{Y} - \bar{X})_{+}^2 + \epsilon_N} \Big\} \approx (\bar{Y} - \bar{X}) \Big\{1 + \frac{(S^2_Y - S^2_X)}{(\bar{Y} - \bar{X})^2} \Big\} = (\bar{Y} - \bar{X}) \ + \ \frac{S^2_Y - S^2_X}{(\bar{Y} - \bar{X})}\equiv g(\bar{X},\bar{Y},S^2_X,S^2_Y).
\end{align*}
Using a first order taylor series expansion we have that
\begin{align*}
		\hat{\delta} \approx g(\mu_X,\mu_Y,\sigma^2_X,\sigma^2_Y) \ &+ \ \frac{\partial g}{\partial \bar{X}} \bigg\rvert_{\bar{X} = \mu_X} (\bar{X} - \mu_X) \ + \ \frac{\partial g}{\partial \bar{Y}} \bigg\rvert_{\bar{Y} = \mu_Y} (\bar{Y} - \mu_Y) \\
			&+ \frac{\partial g}{\partial S^2_X} \bigg\rvert_{S^2_{X} = \sigma^2_X} (S^2_X - \sigma^2_X) + \frac{\partial g}{\partial S^2_Y} \bigg\rvert_{S^2_Y = \sigma^2_Y} (S^2_Y - \sigma^2_Y)
\end{align*}
\begin{align}
	\nonumber	= g(\mu_X,\mu_Y,\sigma^2_X,\sigma^2_Y) &+ \left( \frac{\sigma^2_Y - \sigma^2_X}{(\mu_Y - \mu_X)^2} - 1 \right) (\bar{X} - \mu_X) \ + \ \left( 1 - \frac{\sigma^2_Y - \sigma^2_X}{(\mu_Y - \mu_X)^2} \right) (\bar{Y} - \mu_Y) \\
	&- \ \frac{S^2_X - \sigma^2_X}{(\mu_Y - \mu_X)}  \ + \ \frac{S^2_Y - \sigma^2_Y}{(\mu_Y - \mu_X)}. \label{g}
\end{align}
Now since $\bar{X}, \bar{Y}, S^2_X, S^2_Y$ are all unbiased estimators of $\mu_X, \mu_Y, \sigma^2_X, \sigma^2_Y$ respectively, we have that \\$E[g(\mu_X,\mu_Y,\sigma^2_X,\sigma^2_Y)] \approx \delta$ with accuracy to the first order expansion. Now we derive the variance of $\hat{\delta}$ by taking the variance of (\ref{g})
\begin{align}
	\nonumber Var \left(\hat{\delta} \right) &\approx \left(1 -  \frac{\sigma^2_Y - \sigma^2_X}{(\mu_Y - \mu_X)^2} \right)^2 \cdot \left( \frac{\sigma^2_X}{m} + \frac{\sigma^2_Y}{n} \right) \ + \ 2 \cdot \left( 1 - \frac{\sigma^2_Y - \sigma^2_X}{(\mu_Y - \mu_X)^2} \right) \cdot \left( \frac{\mu_{3cx}/m + \mu_{3cy}/n}{\mu_Y - \mu_X} \right) \\
	 &+ \ \frac{1}{(\mu_Y - \mu_X)^2} \cdot \left( \frac{(\mu_{4cx} - \frac{m-3}{m-1} \sigma^4_X)}{m} + \frac{(\mu_{4cy} - \frac{n-3}{n-1} \sigma^4_Y)}{n} \right).
\end{align}
The special case of $m=n$ gives the asymptotic variance formula in (\ref{sig2d}) as desired.

\newpage

\section*{References}

1. Lindsay, B. G.Mixture models: theory, geometry and applications. In \textit{NSF-CBMS regional conferenceseries in probability and statistics} (1995), JSTOR, pp. i–163. \\
2. McLachlan, G., and Peel, D.John wiley \& sons; new york:  2004. \textit{Finite Mixture Models.} \\
3. FDA, U. Paving  the  way  for  personalized  medicine. \textit{FDA’s  Role  in  a  new  Era  of  Medical  Product Development. US Department of Health and Human Services} (2013), 1–61. \\
4. Spear, B. B., Health-Chiozzi, M., and Huff, J. Clinical application of pharmacogenetics. \textit{Trend in molecular medicine 7}, 5 (2001), 201-204. \\
5. Manegold, C., Adjei, A., Bussolino, F., Cappuzzo, F., Crino, L., Dziadziuszko, R., Et-tinger, D., Fennell, D., Kerr, K., Le Chevalier, T., et al. Novel active agents in patients with advanced nsclc without driver mutations who have progressed after first-line chemotherapy. \textit{ESMO open 1}, 6 (2016), e000118. \\
6. Rosenblatt, J. D., and Benjamini, Y. On mixture alternatives and wilcoxon’s signed-rank test. \textit{The American Statistician 72}, 4 (2018), 344–347. \\
7. Jeske, D. R., and Yao, W. Sample size calculations for mixture alternatives in a control group vs. treatment group design. \textit{Statistics 54}, 1 (2020), 97–113. \\
8. Efron, B. Better bootstrap confidence intervals. \textit{Journal  of  the  American  statistical  Association  82} ,397 (1987), 171–185. \\
9. Singh, Kesar, and Minge Xie. "Bootstrap: a statistical method." \textit{Unpublished manuscript, Rutgers University, USA. Retrieved from http://www.stat.rutgers.edu/home/mxie/RCPapers/bootstrap.pdf} (2008): 1-14. \\
10. Knoll, M. D., and Wonodi, C.Oxford–astrazeneca covid-19 vaccine efficacy. \textit{The Lancet 397}, 10269 (2021), 72–74. \\
11. Ma, Y., Wang, S., Xu, L., and Yao, W. Semiparametric mixture regression with unspecified error distributions. \textit{Test} (2020), 1–16. \\
12. Cho, E., Cho, M. J., and Eltinge, J. The variance of sample variance from a finite population. \textit{International Journal of Pure and Applied Mathematics 21}, 3 (2005), 389. \\
13. Zhang, L. Sample  mean  and  sample  variance: Their  covariance and their (in)dependence. \textit{The American Statistician 61}, 2 (2007), 159–160. \\
14. Dodge, Y., and Rousson, V. The complications of the fourth central moment. \textit{The American Statistician 53}, 3 (1999), 267–269. \\

\end{document}